\documentclass{emulateapj}
\usepackage{apjfonts}

\journalinfo{The Astrophysical Journal Letters, in press}
\slugcomment{Received 2007 June 29; accepted 2007 July 24}

\shortauthors{CAMILO ET AL.}
\shorttitle{THE RADIO-EMITTING MAGNETAR 1E~1547.0--5408}

\begin{document}


\def\mag{1E~1547.0--5408}
\def\psr{PSR~J1550--5418}
\def\snr{G327.24--0.13}
\def\xte{XTE~J1810--197}
\def\asca{{\em ASCA\/}}
\def\chandra{{\em Chandra\/}}
\def\einstein{{\em Einstein\/}}
\def\swift{{\em Swift\/}}
\def\xmm{{\em XMM-Newton\/}}

\title{1E~1547.0--5408: a radio-emitting magnetar with a rotation period
of 2~seconds}

\author{F.~Camilo,\altaffilmark{1}
  S.~M.~Ransom,\altaffilmark{2}
  J.~P.~Halpern,\altaffilmark{1}
  and J.~Reynolds\altaffilmark{3}
}

\altaffiltext{1}{Columbia Astrophysics Laboratory, Columbia University,
  New York, NY 10027.}
\altaffiltext{2}{National Radio Astronomy Observatory, Charlottesville, 
  VA 22903.}
\altaffiltext{3}{Australia Telescope National Facility, CSIRO, Parkes
  Observatory, Parkes, NSW 2870, Australia.}

\begin{abstract}
The variable X-ray source \mag\ was identified by \citet{gg07} as
a likely magnetar in \snr, an apparent supernova remnant.  No X-ray
pulsations have been detected from it.  Using the Parkes radio telescope,
we discovered pulsations with period $P = 2.069$\,s.  Using the Australia
Telescope Compact Array, we localized these to \mag.  We measure $\dot P =
(2.318\pm0.005) \times 10^{-11}$, which for a magnetic dipole rotating
in vacuo gives a surface field strength of $2.2 \times 10^{14}$\,G, a
characteristic age of 1.4\,kyr, and a spin-down luminosity of $1.0 \times
10^{35}$\,ergs\,s$^{-1}$.  Together with its X-ray characteristics,
these rotational parameters of \mag\ prove that it is a magnetar,
only the second known to emit radio waves.  The distance is $\approx
9$\,kpc, derived from the dispersion measure of 830\,cm$^{-3}$\,pc.
The pulse profile at a frequency of 1.4\,GHz is extremely broad and
asymmetric due to multipath propagation in the ISM, as a result of
which only $\approx 75\%$ of the total flux at 1.4\,GHz is pulsed.
At higher frequencies the profile is more symmetric and has $\mbox{FWHM}
= 0.12\,P$.  Unlike in normal radio pulsars, but in common with the other
known radio-emitting magnetar, \xte, the spectrum over 1.4--6.6\,GHz is
flat or rising, and we observe large, sudden changes in the pulse shape.
In a contemporaneous \swift\ X-ray observation, \mag\ was detected
with record high flux, $f_X(1$--$8\,\mbox{keV}) \approx 5 \times
10^{-12}$\,ergs\,cm$^{-2}$\,s$^{-1}$, 16 times the historic minimum.
The pulsar was undetected in archival radio observations from 1998,
implying a flux $< 0.2$ times the present level.  Together with the
transient behavior of \xte, these results suggest that radio emission
is triggered by X-ray outbursts of usually quiescent magnetars.

\end{abstract}


\section{Introduction}\label{sec:intro} 

Anomalous X-ray pulsars (AXPs) and soft gamma-ray repeaters (SGRs)
are young neutron stars with rotation periods of 5--12\,s and inferred
surface magnetic field strength $B \approx 10^{14-15}$\,G (see Woods
\& Thompson 2006 for a review)\nocite{wt06}.  In the magnetar model
\citep[][1996]{dt92a,td95}\nocite{td96a}, the rearrangement and decay
of their extreme fields is responsible for the large and variable X-ray
luminosity of these neutron stars, which exceeds that available from the
braking of their rotation.  Twelve magnetars are confirmed\footnote{Eight
AXPs and four SGRs; there are two more candidates.  See catalog at
http://www.physics.mcgill.ca/$\sim$pulsar/magnetar/main.html.}, of which
two are in supernova remnants (SNRs).  Only one magnetar, the transient
5.5\,s AXP \xte, is known to emit radio waves \citep{crh+06}, with
several unusual characteristics that are not understood.  It is therefore
important to identify further examples of this still mystery-ridden
class of neutron stars.

Discovered with the \einstein\ X-ray satellite in 1980, \mag\ was recently
identified as a magnetar candidate in the center of the small candidate
SNR \snr\ by \citet{gg07}, who argue convincingly against other possible
classifications.  Its flux, as observed by \asca, \chandra, and \xmm,
has varied by a factor of 7, and it has the spectral characteristics
of an AXP.  Although no X-ray pulsations have been detected from \mag,
the upper limit of 14\% on its pulsed fraction is larger than that of
one known AXP, 4U~0142+62.  Here we report the discovery and initial
study of radio pulsations from \mag, confirming that it is a magnetar.

\section{Observations, Analysis, and Results}\label{sec:obs} 

\subsection{Radio Pulsar Discovery} \label{sec:parkes}

We observed \mag\ with the Parkes telescope in Australia
on 2007 June 8.  We collected data for 20 minutes using the
central beam of the multibeam receiver at a frequency $\nu =
1.374$\,GHz.  A bandwidth of 288\,MHz was recorded, divided into
96 frequency channels sampled every 1\,ms with one-bit precision.
We analyzed the data using standard techniques implemented in
PRESTO\footnote{http://www.cv.nrao.edu/$\sim$sransom/presto/.}
\citep{rem02}, in similar fashion to other pulsar searches
\citep[e.g.,][]{crg+06}.  We detected a periodic signal with $P=2.069$\,s
that is dispersed ($\mbox{DM} \approx 830$\,cm$^{-3}$\,pc) and hence of
astronomical origin, which we name \psr.

\begin{figure}[ht]
\begin{center}
\includegraphics[angle=0,scale=0.55]{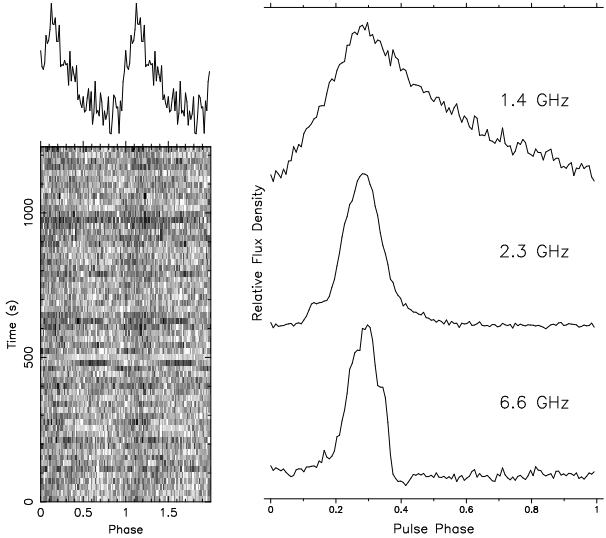}
\caption{\label{fig:profs}
Detections of \psr\ (\mag) at Parkes, with $P = 2.069$\,s. {\em Left}:
Discovery observation on 2007 June 8 at 1.4\,GHz.  The profile is shown
twice (summed at the top).  {\em Right}:  Average profiles at 1.4, 2.3,
and 6.6\,GHz, based on 7.6, 1.8, and 0.4\,hr of data, respectively.
At 6.6 and 2.3\,GHz, the profiles have $\mbox{FWHM} = 0.12\,P$ and
$0.14\,P$, respectively, while the huge asymmetric tail of the 1.4\,GHz
profile is caused by scattering in the ISM.  Each profile was corrected
for distortions introduced by a high-pass filter in the data path with
a time constant of 0.9\,s \citep[see][]{mlc+01,crh+06}.  The profiles
were aligned by eye.  }
\end{center}
\end{figure}

\begin{figure}[ht]
\begin{center}
\includegraphics[angle=0,scale=0.43]{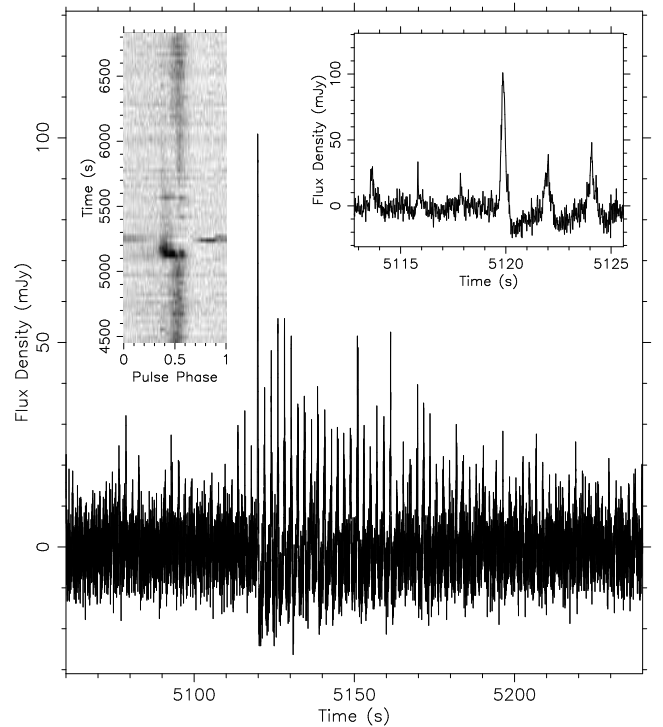}
\caption{\label{fig:sp}
Individual pulses and changing profiles from the 2-s \psr\ at a
frequency of 2.3\,GHz observed at Parkes on 2007 June 20. {\em Main}:
A time series of 87 pulses containing the brightest pulse visible in the
data set.  At approximately this time (darkest 30\,s sub-integration;
{\em left inset}), the average profile changed significantly, with the low
shoulder visible in Fig.~\ref{fig:profs} becoming temporarily brighter,
and ``recovering'' after about 20 minutes.  In the left inset panel,
a burst of radio-frequency interference is visible 2 minutes after the
largest pulse.  {\em Right inset}:  A zoom-in on the main panel, showing
the structure of six single pulses.  In both panels the brightest pulse
($\mbox{FWHM} \approx 0.11\,P$) is the same, and the data are displayed
with 16\,ms resolution.  The dip following each strong pulse is an
artifact.  }
\end{center}
\end{figure}

We observed the pulsar on 11 occasions spanning 19 days, detecting it
every time.  The pulse profile at 1.4\,GHz is very broad and asymmetric,
dominated by multipath propagation in the ISM (Fig.~\ref{fig:profs}).
The calibrated period-averaged pulsed flux density is $S_{1.4}
= 2.5\pm0.5$\,mJy, with daily-averaged fluxes constant within the
uncertainties.  However, because the pulse is so scattered at 1.4\,GHz,
about 25\% of the received flux is unpulsed, which we measure with
the Australia Telescope Compact Array (ATCA; \S~\ref{sec:atca}).
At higher frequencies the profile is more symmetric and narrower
(Fig.~\ref{fig:profs}), and the flux density is larger ($S_{6.6} \approx
6$\,mJy in our only observation at 6.6\,GHz), suggesting a rising
spectrum.  In one of two observations at 2.3\,GHz we saw sudden, major
profile variations (see Fig.~\ref{fig:sp}), and until they are better
characterized, spectral indices should be estimated from simultaneous
multi-frequency measurements.  We measured pulse arrival times for
every observation and obtained a phase-connected timing solution with
TEMPO\footnote{http://www.atnf.csiro.au/research/pulsar/tempo/.}.
The parameters from this fit are listed in Table~\ref{tab:parms}.

The position of \psr\ was observed during the Parkes multibeam Galactic
plane survey \citep[e.g.,][]{mlc+01}.  The closest pointings were made on
1998 August 8 and 13, offset by $3.1'$ (survey ID 4859499).  We searched
for pulsations in these data, but found none to a limit of $\la 0.5$\,mJy
(accounting for the less sensitive beam used at an offset position).
Since interstellar scintillation is not expected to modulate the flux of
a pulsar with its DM \citep{ssh+00}, these non-detections imply that the
pulsar was fainter in 1998 August than in 2007 June by a factor $\ga 5$.

At 2\,s, the rotation period of \psr\ is substantially smaller than
that of any other known magnetar.  Nevertheless, its $\dot P$ implies a
magnetic field solidly in the range of magnetars ($B \equiv 3.2 \times
10^{19}(P \dot P)^{1/2}\,\mbox{G} = 2 \times 10^{14}$\,G).  The $\mbox{DM}
= 830$\,cm$^{-3}$\,pc is consistent with the large X-ray-fitted neutral
hydrogen column density, $N_{\rm H} \approx 3 \times 10^{22}$\,cm$^{-2}$
\citep{gg07}, and implies a distance $d \approx 9$\,kpc according to the
free-electron model of \citet{cl02}.  A smaller $d \approx 4$\,kpc for
the candidate SNR~\snr\ was suggested by \citet{gg07}.  Assuming that
the two objects are associated, the smaller distance estimate would
place the magnetar/SNR in or near the Crux--Scutum spiral arm, while
the larger distance would be compatible with a Norma spiral arm location.

The DM model also predicts a $\sim 10$\,ms scattering timescale
at 1.4\,GHz.  Instead, the $\sim 1$\,s broadening observed
(Fig.~\ref{fig:profs}) is larger than for any known pulsar
\cite[see][]{bcc+04}.  If a portion of the unmodeled scattering were
caused by electrons with larger average density in this direction
than contained in the model, then the predicted distance could be an
overestimate.  The 2.3\,GHz profile shown in Figure~\ref{fig:profs},
while much more symmetric, is still scattered, by a fitted amount $\approx
100$\,ms that is consistent with the approximate $\nu^{-4.4}$ scaling.

\subsection{Contemporaneous Radio Imaging} \label{sec:atca}

Identification of \psr\ with \mag\ can be ensured by measuring its radio
position with high precision, as the X-ray position is already known to
$0.8''$ from \chandra\ \citep{gg07}.  With this aim, on 2007 June 26 we
made a pulsar-gated observation at the ATCA.  \mag\ was observed at 1.384
and 2.368\,GHz simultaneously.  The array was in its 6C configuration,
with antennas CA01 and CA05 off line.  At each frequency, we sampled a
bandwidth of 128\,MHz split into 33 channels for each of two orthogonal
linear polarizations.  Visibilities were accumulated into 32 phase bins
coherently with the pulsar period.  First we observed the flux calibrator
0823--500, and then obtained data for 7\,hr over a 12\,hr span, with
scans of 20 minutes on the pulsar interleaved with 90\,s scans on the
phase calibrator 1613--586.  The configuration used was, in effect,
only sensitive to point sources.

We analyzed the correlated data using standard techniques within
MIRIAD\footnote{http://www.atnf.csiro.au/computing/software/miriad/.}.
We detected pulsations from \psr\ at both frequencies at a position
identical to \chandra's to within $0.1''$ in each coordinate,
greatly reducing the uncertainty (see Table~\ref{tab:parms}).
The pulsar phase-integrated flux densities measured on this day were
$S_{1.4}=3.3\pm0.3$\,mJy and $S_{2.4}=5.3\pm0.3$\,mJy.  The corresponding
spectral index is $\alpha=0.9\pm0.2$ ($S_{\nu} \propto \nu^{\alpha}$).
The one $S_{6.6} \approx 6$\,mJy measurement (\S~\ref{sec:parkes})
suggests that this rising spectrum may flatten at higher frequencies,
but due to the potential for variability, further measurements will be
needed to address this question.  We have not detected any flux from the
off-pulse phase bins at 2.4\,GHz, from which we limit any emission from
a nebula smaller than the $3''$ beam size to $<0.7$\,mJy ($3\,\sigma$).

\subsection{Contemporaneous X-ray Observation}

On 2007 June 22 we obtained a 2941\,s exposure of \mag\ with the \swift\
X-ray telescope in photon counting mode, which provides imaging with
$18''$ half-power diameter resolution and 2.5\,s time sampling.
The source was detected in the range 1--8\,keV with a mean count
rate of 0.095\,s$^{-1}$, with a spectrum similar to that from \xmm\
observations \citep{gg07}.  Neither the counts nor the time resolution
were sufficient to detect pulsations.  However, the flux was $\approx 5
\times 10^{-12}$\,ergs\,cm$^{-2}$\,s$^{-1}$, which is 16 times greater
than the minimum seen in 2006 July--August, and 2.4 times greater than
the previous highest flux.   The bolometric blackbody flux corrected for
absorption is $\approx 2 \times 10^{-11}$\,ergs\,cm$^{-2}$\,s$^{-1}$.
Further details of this and subsequent X-ray observations will be
reported separately.

\begin{deluxetable}{ll}
\tablewidth{0.88\linewidth}
\tablecaption{\label{tab:parms} Timing Parameters of \psr\ }
\tablecolumns{2}
\tablehead{
\colhead{Parameter} &
\colhead{Value}
}
\startdata
R.A.\tablenotemark{a} (J2000)\dotfill  & $15^{\rm h}50^{\rm m}54.11^{\rm s}\pm0.01^{\rm s}$ \\
Decl.\tablenotemark{a} (J2000)\dotfill & $-54\arcdeg18'23.7''\pm0.1''$        \\
Epoch (MJD)\dotfill                              & 54270.0                    \\
Spin period, $P$ (s)\dotfill                     & 2.06983302(4)              \\
Period derivative, $\dot P$\dotfill              & $2.318(5) \times 10^{-11}$ \\
Dispersion measure\tablenotemark{b}, DM (cm$^{-3}$\,pc)\dotfill & $830\pm50$  \\
Timing fit rms residual ($P$)\dotfill            & 0.013                      \\
Range of timing solution (MJD)\dotfill           & 54259--54278               \\
Surface magnetic field, $B$ (G)\dotfill                 & $2.2 \times 10^{14}$\\
Characteristic age, $\tau_c$ (kyr)\dotfill              & 1.4                 \\
Spin-down luminosity, $\dot E$ (ergs\,s$^{-1}$)\dotfill & $1.0 \times 10^{35}$\\
Distance\tablenotemark{c}, $d$ (kpc)\dotfill            & $\approx 9$         
\enddata
\tablecomments{Numbers in parentheses are the TEMPO 1\,$\sigma$
uncertainties. }
\tablenotetext{a}{Coordinates from the 2.4\,GHz ATCA image
(\S~\ref{sec:atca}). }
\tablenotetext{b}{The large DM uncertainty is due to uncertain
frequency-dependent profile fiducial points caused by scattering. }
\tablenotetext{c}{Obtained from the DM with the \citet{cl02} model. }
\end{deluxetable}

\section{Discussion}\label{sec:disc} 

The rotational parameters of \psr, the X-ray spectrum and variability
of \mag, and the positional coincidence between them, prove that they
are the same object --- a radio-emitting magnetar.

\subsection{General Properties in Relation to Other Magnetars}

The spin-down luminosity of \psr, $\dot E \equiv 4\pi^2 I \dot P P^{-3}
= 1 \times 10^{35}$\,ergs\,s$^{-1}$, exceeds that of any previously
known magnetar.  Typical values of $\dot E$ for AXPs are $\sim 1
\times 10^{33}$\,ergs\,s$^{-1}$, while they range from $6 \times
10^{31}$\,ergs\,s$^{-1}$ for 1E~2259+586 \citep{kno+89,bs96} to the
historic maximum of $5 \times 10^{34}$\,ergs\,s$^{-1}$ for SGR~1806--20
\citep{mte+05}.  The spin-down luminosity of \psr\ is only slightly
exceeded by its maximum X-ray luminosity, $L_X(0.5$--$10\,\mbox{keV})
\approx 1.9 \times 10^{35}\ (d/9\,\mbox{kpc})^2$\,ergs\,s$^{-1}$, which is
one of the discriminators of a magnetar from a rotation-powered pulsar.
It is unlikely that a significant fraction of this X-ray luminosity is
powered by rotation; its distinctive properties are consistent with
crustal heating by magnetic field decay as in other AXPs.  First,
the luminosity has varied by a factor of 16, while rotation-powered
pulsars and cooling neutron stars are steady.  Second, the composite
X-ray spectrum is similar to that of other AXPs, including a thermal
component that even in a low state is fitted by a blackbody of $kT =
0.43$\,keV \citep{gg07}, which is hotter by at least a factor of 3 than
a cooling neutron star of its same age \citep[e.g.,][]{ykhg02}.

That \psr\ has the shortest $P$ and highest $\dot E$ among known magnetars
is consistent with the hypothesis that they are born with $P \ll 1$\,s and
spin down to their longest observed periods on timescales of $10^4$\,yr.
However, it does not ease the problem of the narrow period distribution of
magnetars \citep{pm02}, in particular, why there is a cutoff at long $P$.

\subsection{Birth Properties of \mag }

The small characteristic age of \psr\ may be related to the size
of its presumed SNR~\snr, which has a diameter of $4' = 4.7\,(d/
4\,\mbox{kpc})$\,pc \citep{gg07}.  We first argue that the SNR hosts
of AXPs are no older than their pulsar's characteristic age: the two
previously confirmed AXPs that are located in remnants have characteristic
age slightly larger than the SNR age \citep[1E~1841--045;][]{vg97},
or an order of magnitude larger \citep[1E~2259+586;][]{wql+92,rp97}.
In the latter case, it is possible that the historic average of $\dot P$
was much higher than its present value.  Similarly, \snr\ is probably
not older than the $\tau_c \equiv P/2\dot P = 1.4$\,kyr of \psr.  Then,
at $d = 4$\,kpc, the radius of the SNR shell implies an average expansion
velocity of 1600\,km\,s$^{-1}$.  This is small for the free-expansion
stage of an SNR.  Assuming instead that the SNR is already in the
adiabatic (Sedov) phase, and following \citet{vk06} who argued that the
explosion energies of AXP remnants are no larger than those of ordinary
pulsars, we find, for an explosion energy of $10^{51}$\,ergs, an ambient
density $\rho = 2.1 \times 10^{-22}$\,g\,cm$^{-3}$, or $n_{\rm H} =
90$\,cm$^{-3}$, which is typical of molecular clouds.  If the distance
is 9\,kpc, then the required density is $n_{\rm H} = 1.5$\,cm$^{-3}$.
These densities are compatible with the expectation that magnetars are
born from massive stars that have short lives and are the first ones
to explode, in environments that are still dense \citep[see][]{gmo+05}.
The larger distance, however, requires less extreme conditions than the
nearer one.

The location of \mag\ in the center of its presumed SNR, similar to
1E~2259+586 and 1E~1841--045, is an indication that magnetars are not born
with greater kick velocity than normal pulsars.  The only direct proper
motion measurement of a magnetar, \xte\ with tangential velocity $V_\perp
= 212$\,km\,s$^{-1}$ \citep{hcb+07}, supports this interpretation.
We judge by eye that \mag\ is within $30''$ of the center of the $4'$
diameter shell of \snr\ \citep[Fig.~4 of][]{gg07}, its presumed birth
location.  If its true age is equal to or greater than its characteristic
age of 1.4\,kyr, then its proper motion is less than $0.02''$\,yr$^{-1}$.
This corresponds to $V_\perp < 900\ (d/9\,\mbox{kpc})$\,km\,s$^{-1}$.

\subsection{Comparison with \xte }

Two AXPs, \mag\ and \xte, are now known to emit radio waves, and a
comparison of their properties is revealing.  Both appear to be transient.
In the case of \xte, the radio emission began within 1\,yr of its only
known X-ray outburst \citep{hgb+05}.  At its observed peak $>3$\,yr
after the X-ray outburst, the radio flux density was $>50$ times the
prior upper limit \citep{crh+06}, but the X-rays have now returned to
quiescence, $>100$ times below their peak flux \citep{gh07}, and the
radio flux is much diminished as well \citep{ccr+07}.  For \psr, we know
that in 2007 June its radio flux density is $>5$ times the upper limit
from 1998 August, while the contemporaneous X-ray flux is the highest
observed.  The X-ray flux history of \mag\ was not as well sampled nor as
consistently faint as that of \xte, but it has varied by a factor of 16.

It may be significant that the two AXPs detected so far in radio have the
smallest periods, 2.0 and 5.5\,s.  While X-ray emission from magnetars is
manifestly not powered by rotation, it is not ruled out that their radio
emission is governed by the same polar-cap gap accelerators and death
lines as ordinary pulsars.  If coming from open field lines, the width of
the radio beam is proportional to $P^{-1/2}$, which favors detection of
short-period pulsars if all are radio emitters and are randomly aligned.
On the other hand, the 6.6\,GHz pulse from \psr\ (Fig.~\ref{fig:profs})
is much wider than almost all profiles of ordinary long-period pulsars,
so a different explanation may have to be sought for its pulse width.

If the viewing angle and magnetic inclination angle with respect to
the rotation axis are both small, this could explain the broad radio
pulse of \psr\ and the failure so far to see pulsed X-ray modulation
(assumed to come from surface thermal emission).  It is also possible
that the wide profile of \psr\ is indicative of emission from closed,
non-potential field lines in the magnetar model \citep{tlk02} instead
of the narrow open field-line bundle.  Closed, twisted field lines may
span a large range of azimuthal angles near the surface of the neutron
star, and models of pair production there yield relativistic $\gamma
\sim 10^3$ that are appropriate for radio emission \citep{bt07}.
From \xte, there are indications that at least some of the radio
emission originates on open field lines \citep{ccr+07,crj+07}, while some
characteristics remain unexplained and could point to either location
\citep{ccr+07,crp+07,ksj+07}.

Both pulsars have exhibited sudden changes in radio pulse shape.
The example in Figure~\ref{fig:sp} is somewhat reminiscent of X-ray
bursts from AXPs including \xte\ \citep{wkg+05}, although X-ray bursts
were not seen in coincidence with such radio transitions in \xte\
\citep{ccr+07}.  Based on the few observations so far, pulse-shape
variations observed from \psr\ are less pronounced than in \xte, and
its daily-averaged flux density may be steady, unlike the first radio
magnetar \citep{crh+06,ccr+07}.  Most striking are the flat or inverted
radio spectra of \psr\ and \xte\ \citep{crp+07}, which are unique,
and clearly distinguishes them from ordinary radio pulsars.

At their peak, both magnetars are very luminous young radio pulsars:
$L_{1.4} \equiv S_{1.4} d^2 \ga 100$\,mJy\,kpc$^2$, which is larger than
the $L_{1.4}$ of virtually any ordinary pulsar with $\tau_c < 10^4$\,yr
\citep[see][]{cmg+02}.  If this is a general property, the future may be
(transiently) bright for further radio detections from known or yet to
be identified magnetars.

\acknowledgements
We are indebted to S.\ Johnston, C.\ Phillips, G.\ Hobbs, and J.\ Verbiest
for generously giving us some of their observing time, and P.\ Edwards
for quickly approving and scheduling the observation at the ATCA.  R.\
Bhat kindly determined the scattering timescale of the profile.  We thank
M.\ Kramer and M.\ Keith for providing us with archival data from the
Parkes multibeam Galactic plane survey.  We are grateful to the \swift\
project for the prompt approval and scheduling of our observing request,
and to N.\ Mirabal for help with the analysis.  The Parkes Observatory
and the ATCA are part of the Australia Telescope, which is funded by the
Commonwealth of Australia for operation as a National Facility managed
by CSIRO.  This work was supported in part by the NSF through grant
AST-05-07376 to F.C., who also made use of the NRAO travel fund.


\begin{thebibliography}{33}
\expandafter\ifx\csname natexlab\endcsname\relax\def\natexlab#1{#1}\fi

\bibitem[{Baykal \& Swank(1996)}]{bs96}
Baykal, A., \& Swank, J. 1996, ApJ, 460, 470

\bibitem[{Beloborodov \& Thompson(2007)}]{bt07}
Beloborodov, A.~M., \& Thompson, C. 2007, ApJ, 657, 967

\bibitem[{{Bhat} {et~al.}(2004){Bhat}, {Cordes}, {Camilo}, {Nice}, \&
  {Lorimer}}]{bcc+04}
{Bhat}, N.~D.~R., {Cordes}, J.~M., {Camilo}, F., {Nice}, D.~J., \& {Lorimer},
  D.~R. 2004, ApJ, 605, 759

\bibitem[{{Camilo} {et~al.}(2007{\natexlab{a}}){Camilo}, Cognard, {Ransom},
  {Halpern}, {Reynolds}, {Zimmerman}, Gotthelf, Helfand, Demorest, Theureau, \&
  Backer}]{ccr+07}
{Camilo}, F., et al. 2007{\natexlab{a}}, ApJ, 663, 497

\bibitem[{Camilo {et~al.}(2002)Camilo, Manchester, Gaensler, Lorimer, \&
  Sarkissian}]{cmg+02}
Camilo, F., Manchester, R.~N., Gaensler, B.~M., Lorimer, D.~L., \& Sarkissian,
  J. 2002, ApJ, 567, L71

\bibitem[{{Camilo} {et~al.}(2006{\natexlab{a}}){Camilo}, {Ransom}, {Gaensler},
  {Slane}, {Lorimer}, {Reynolds}, {Manchester}, \& {Murray}}]{crg+06}
{Camilo}, F., {Ransom}, S.~M., {Gaensler}, B.~M., {Slane}, P.~O., {Lorimer},
  D.~R., {Reynolds}, J., {Manchester}, R.~N., \& {Murray}, S.~S.
  2006{\natexlab{a}}, ApJ, 637, 456

\bibitem[{{Camilo} {et~al.}(2006{\natexlab{b}}){Camilo}, {Ransom}, {Halpern},
  {Reynolds}, {Helfand}, {Zimmerman}, \& {Sarkissian}}]{crh+06}
{Camilo}, F., {Ransom}, S.~M., {Halpern}, J.~P., {Reynolds}, J., {Helfand},
  D.~J., {Zimmerman}, N., \& {Sarkissian}, J. 2006{\natexlab{b}}, Nature, 442,
  892

\bibitem[{{Camilo} {et~al.}(2007{\natexlab{b}}){Camilo}, Ransom, Pe\~{n}alver,
  Karastergiou, van Kerkwijk, Durant, {Halpern}, Reynolds, Thum, Helfand,
  Zimmerman, \& Cognard}]{crp+07}
{Camilo}, F., et al. 2007{\natexlab{b}}, ApJ, in press (arXiv:0705.4095)

\bibitem[{{Camilo} {et~al.}(2007{\natexlab{c}}){Camilo}, Reynolds, Johnston,
  {Halpern}, Ransom, \& van Straten}]{crj+07}
{Camilo}, F., Reynolds, J., Johnston, S., {Halpern}, J.~P., Ransom, S.~M., \&
  van Straten, W. 2007{\natexlab{c}}, ApJ, 659, L37

\bibitem[{{Cordes} \& {Lazio}(2002)}]{cl02}
{Cordes}, J.~M., \& {Lazio}, T.~J.~W. 2002, preprint (astro-ph/0207156)

\bibitem[{Duncan \& Thompson(1992)}]{dt92a}
Duncan, R.~C., \& Thompson, C. 1992, ApJ, 392, L9

\bibitem[{{Gaensler} {et~al.}(2005){Gaensler}, {McClure-Griffiths}, {Oey},
  {Haverkorn}, {Dickey}, \& {Green}}]{gmo+05}
{Gaensler}, B.~M., {McClure-Griffiths}, N.~M., {Oey}, M.~S., {Haverkorn}, M.,
  {Dickey}, J.~M., \& {Green}, A.~J. 2005, ApJ, 620, L95

\bibitem[{{Gelfand} \& {Gaensler}(2007)}]{gg07}
{Gelfand}, J.~D., \& {Gaensler}, B.~M. 2007, ApJ, in press (arXiv:0706.1054)

\bibitem[{{Gotthelf} \& {Halpern}(2007)}]{gh07}
{Gotthelf}, E.~V., \& {Halpern}, J.~P. 2007, Ap\&SS, 308, 79

\bibitem[{Halpern {et~al.}(2005)Halpern, Gotthelf, Becker, Helfand, \&
  White}]{hgb+05}
Halpern, J.~P., Gotthelf, E.~V., Becker, R.~H., Helfand, D.~J., \& White, R.~L.
  2005, ApJ, 632, L29

\bibitem[{Helfand {et~al.}(2007)Helfand, Chatterjee, Brisken, Camilo, Reynolds,
  van Kerkwijk, Halpern, \& Ransom}]{hcb+07}
Helfand, D.~J., Chatterjee, S., Brisken, W., Camilo, F., Reynolds, J., van
  Kerkwijk, M.~H., Halpern, J.~P., \& Ransom, S.~M. 2007, ApJ, 662, 1198

\bibitem[{{Koyama} {et~al.}(1989){Koyama}, {Nagase}, {Ogawara}, {Shinoda},
  {Kawai}, {Jones}, {Williams}, {Watson}, {Makishima}, \& {Ohashi}}]{kno+89}
{Koyama}, K., et al. 1989, \pasj, 41, 461

\bibitem[{{Kramer} {et~al.}(2007){Kramer}, {Stappers}, {Jessner}, {Lyne}, \&
  {Jordan}}]{ksj+07}
{Kramer}, M., {Stappers}, B.~W., {Jessner}, A., {Lyne}, A.~G., \& {Jordan},
  C.~A. 2007, MNRAS, 377, 107

\bibitem[{Manchester {et~al.}(2001)Manchester, Lyne, Camilo, Bell, Kaspi,
  D'Amico, McKay, Crawford, Stairs, Possenti, Morris, \& Sheppard}]{mlc+01}
Manchester, R.~N., et al. 2001, MNRAS, 328, 17

\bibitem[{{Mereghetti} {et~al.}(2005){Mereghetti}, {Tiengo}, {Esposito},
  {G{\"o}tz}, {Stella}, {Israel}, {Rea}, {Feroci}, {Turolla}, \&
  {Zane}}]{mte+05}
{Mereghetti}, S., et al. 2005, \apj, 628, 938

\bibitem[{{Psaltis} \& {Miller}(2002)}]{pm02}
{Psaltis}, D., \& {Miller}, M.~C. 2002, \apj, 578, 325

\bibitem[{{Ransom} {et~al.}(2002){Ransom}, {Eikenberry}, \&
  {Middleditch}}]{rem02}
{Ransom}, S.~M., {Eikenberry}, S.~S., \& {Middleditch}, J. 2002, AJ, 124, 1788

\bibitem[{Rho \& Petre(1997)}]{rp97}
Rho, J., \& Petre, R. 1997, ApJ, 484, 828

\bibitem[{Stinebring {et~al.}(2000)Stinebring, Smirnova, Hankins, Hovis, Kaspi,
  Kempner, Meyers, \& Nice}]{ssh+00}
Stinebring, D.~R., Smirnova, T.~V., Hankins, T.~H., Hovis, J., Kaspi, V.,
  Kempner, J., Meyers, E., \& Nice, D.~J. 2000, ApJ, 539, 300

\bibitem[{{Thompson} \& Duncan(1995)}]{td95}
Thompson, C., \& {Duncan}, R.~C. 1995, MNRAS, 275, 255

\bibitem[{Thompson \& Duncan(1996)}]{td96a}
Thompson, C., \& Duncan, R.~C. 1996, ApJ, 473, 322

\bibitem[{Thompson {et~al.}(2002)Thompson, Lyutikov, \& Kulkarni}]{tlk02}
Thompson, C., Lyutikov, M., \& Kulkarni, S.~R. 2002, ApJ, 574, 332

\bibitem[{Vasisht \& Gotthelf(1997)}]{vg97}
Vasisht, G., \& Gotthelf, E.~V. 1997, ApJ, 486, L129

\bibitem[{Vink \& Kuiper(2006)}]{vk06}
Vink, J., \& Kuiper, L. 2006, MNRAS, 370, L14

\bibitem[{{Wang} {et~al.}(1992){Wang}, {Qu}, {Luo}, {McCray}, \& {Mac
  Low}}]{wql+92}
{Wang}, Z., {Qu}, Q., {Luo}, D., {McCray}, R., \& {Mac Low}, M.-M. 1992, ApJ,
  388, 127

\bibitem[{Woods \& Thompson(2006)}]{wt06}
Woods, P.~M., \& Thompson, C. 2006, in Compact Stellar X-ray Sources, ed.
  W.~H.~G. Lewin \& M.~van~der Klis (Cambridge: Cambridge Univ. Press),
  547

\bibitem[{{Woods} {et~al.}(2005){Woods}, {Kouveliotou}, {Gavriil}, {Kaspi},
  {Roberts}, {Ibrahim}, {Markwardt}, {Swank}, \& {Finger}}]{wkg+05}
{Woods}, P.~M., et al. 2005, \apj, 629, 985

\bibitem[{{Yakovlev} {et~al.}(2002){Yakovlev}, {Kaminker}, {Haensel}, \&
  {Gnedin}}]{ykhg02}
{Yakovlev}, D.~G., {Kaminker}, A.~D., {Haensel}, P., \& {Gnedin}, O.~Y. 2002,
  A\&A, 389, L24

\end{thebibliography}
\end{document}